# A Motivation Model of Peer Assessment in Programming Language Learning

**Yanqing Wang, Yaowen Liang, Luning Liu**
School of Management, Harbin Institute of Technology, Harbin 150001, China
yanqing@hit.edu.cn // hitliangyaowen@hotmail.com // liuluning@hit.edu.cn

**Ying Liu**
Department of Information Systems, College of Business Administration, California State University Long Beach, CA 90840, USA
Ying.Liu@csulb.edu

**ABSTRACT**

Peer assessment is an efficient and effective learning assessment method that has been used widely in diverse fields in higher education. Despite its many benefits, a fundamental problem in peer assessment is that participants lack the motivation to assess others' work faithfully and fairly. Non-consensus is a common challenge that makes the reliability of peer assessment a primary concern in practices. This research proposes a motivation model that uses review deviation and radicalness to identify non-consensus in peer assessment. The proposed model is implemented as a software module in a peer code review system called EduPCR4. EduPCR4 is able to monitor this measure and trigger teacher's arbitration when it detects possible non-consensus. An empirical study conducted in a university-level C programming course showed that the proposed model and its implementation helped to improve the peer assessment practices in many aspects.

## Introduction

Peer assessment research has a long history dating back to 1920s (Kane & Lawler, 1978). It is an educational arrangement where students judge peers' performance quantitatively and/or qualitatively (Zundert et al., 2010). It stimulates students to reflect, discuss, and collaborate in their learning process (Topping, 1998; Strijbos & Sluijsmans, 2010). Peer assessment is recommended because it reduces faculty's workload (Rubin & Turner, 2012) and increases learning outcome (Murakami et al., 2012). Due to its efficiency and active learning nature, peer assessment has been widely used in diverse fields (Falchikov, 1995; Freeman, 1995). In a recent study on undergraduate peer assessments, students reported positive experience in peer assessment and recommended its use in college courses (Vickerman, 2009).

The effectiveness and quality of an assessment depends on how it is incorporated into the learning process (Segers et al., 2003). Researchers have argued that peer assessment is a rather complex undertaking because of the close relationship of content knowledge and assessment quality (Sluijsmans et al., 2002). Students may not have enough content knowledge to criticize a peer's work and conduct a fair evaluation. To alleviate this problem, Tseng and Tsai (2007) developed a peer assessment process that consists of three rounds of reviews. Because each round has a different reviewer, the bias introduced by a single reviewer is significantly reduced. It is an improvement by assigning multiple reviewers to an evaluation task. Moreover, Chen et al. (2009) explored how high level prompts and peer assessment can affect a learner's reflection levels in an online learning context. Lin et al. (2011) investigated the impact of an online reflective peer assessment on students' argumentation and conceptual understanding. For all that, there are still many other issues in peer assessment. In any assessment task, a reviewer's motivation plays an important role in the assessment quality. Rather less attention has been paid to improve student motivation in peer assessment.

Peer assessment has been used in different computer programming courses for many years (Li, 2007; Wang et al., 2008). In those courses, each student acts as both an author and a reviewer. There are positive feedbacks including more efficient and more effective learning experience (Li & Joyce, 2008; Wang et al., 2012). However, issues such as the lack of motivation (Higgins et al., 2002; Papinczak et al., 2007), non-consensus (Chang et al., 2011) raise concerns about the reliability of peer assessment. Teachers have to spend lots of time on inspecting and assessing the programs of every student to solve those issues.

To address the reliability issue, we propose a motivation model that incorporates student motivation, non-consensus measure. The model is implemented in the fourth edition of an *Educational Peer Code Review* (EduPCR4) information system, which supports multiple-reviewer assignment. More importantly, EduPCR4 implements a strategy that motivates students' participation in peer assessment activities. The system automatically calculates and monitors the reliability of the peer assessment results based on the motivation model measures. The remainder of this paper is organized as follows. After the motivation model is formally defined, we discuss the challenge to the motivation model and its solution. Then we describe the empirical study of the proposed model. We give conclusions and future research directions finally.

## Motivation Model

### Definitions

The terms and concepts used in the proposed motivation model are listed in Table 1.

*Table 1.* Definitions and their meanings in this paper

| Definition | Meaning |
|---|---|
| role | There are three roles in this model: *author*, *reviewer*, and *teacher*. Every student plays both the author role and the reviewer role in each task. |
| task | One task is a programming assignment that has several steps including source code submission, code review, code revision, and reverse review. Students should complete each step before a given deadline. |
| source code | The program source code written by a student as an author. The source code should be successfully compiled, built, and tested by its author before submission. |
| code review | A code review includes literal comments and a numerical score. Both are given by a peer reviewer as evaluation results to another student's source code. |
| revision code | A new edition of program source code submitted by its original author based on the review comments. |
| reverse review | It is the review done by an author to evaluate the code review given by a peer reviewer. The author reads the source code comments and gives a score to the code review. The score is based on quality measures such as *constructiveness*, *informativeness*, and *fairness* of the comments. |
| document | An electronic document contains one of four types of programming artifacts: *source code*, *code review*, *revision code*, and *reverse review*. Code review and reverse review are stored in database. Source code and revision code are saved as files. |
| process | A teacher creates a task in EduPCR4 and specifies deadlines for all documents mentioned in the previous row. Each student first completes source code and submits it to EduPCR4. A group of designated reviewers reviews and assesses the source code online. The author examines the code review, gives reverse review, revises source code, and submits revision code. Finally, EduPCR4 calculates the overall score of each student in a task according to the rules depicted in Appendix. |
| multiple reviewers | In a task, a review group that consists of multiple reviewers is assigned to an author's source code in order to make peer assessment results more reliable. |
| non-consensus | To an identical source code, reviewers have quite different options and consequently give quite different scores. It is a normal phenomenon in multiple-reviewer peer assessment process. |
| teacher arbitration | When a group of reviewers fails to reach a consensus, EduPCR4 sends the class teacher a notice. The teacher should examine all review comments and scores to figure out the reasons of discrepancy and take necessary actions to fix any unfair scores. The teacher has the right to override any review score. |

**Motivation Model Activities**

There are a number of activities performed by different roles in each task. The roles and activities are depicted in Fig. 1. An activity may generate scores. There are three types of scores including *process score*, *quality score*, and *bonus score*. Their weights can be defined by the teacher in charge of a course. In the activity diagram shown in Fig. 1, *codeScore* and *reviewScore* belong to quality score; *sourceDone*, *revisionDone*, *reviewDone*, and *reverseDone* are process scores; *reviewBonus* is the only bonus score. In a multiple-reviewer task, except for *sourceDone* and *revisionDone*, each score item in Fig. 1 is an array (*integer* or *boolean* type). For example, if a review group has five peer reviewers, each score item has five values.

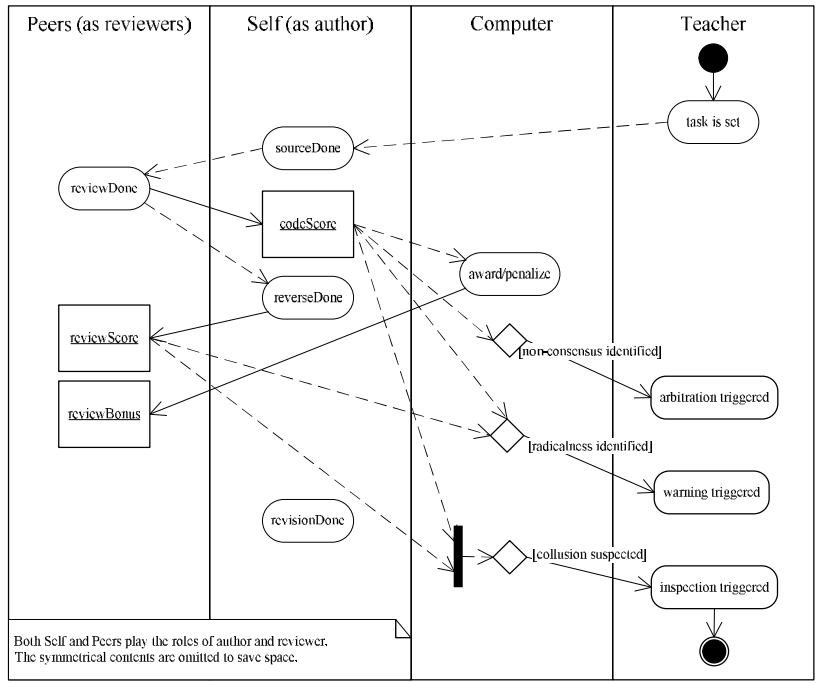

*Figure 1.* Activity diagram of the motivational model

The three types of scores are described in detail as follows.
- *Process scores* are used to motivate students to submit task documents in time. When a student submits a document before deadline, he or she will get a fixed-value score. Late submission will have a score of zero. The process score includes three parts: whether source code and revision code are submitted in time (*sourceDone* and *revisionDone* in Fig. 2), whether an assigned review work is done in time (*reviewDone* in Fig. 2), and whether a reverse review is done in time (*reverseDone* in Fig. 2).
- *Quality scores* are used to measure the quality of source code and code review comments. A source code score (*codeScore* in Fig. 2) is the mean value of all *codeScore* values given by a group of peer reviewers. Similarly, review quality score (*reviewScore* in Fig. 2) is the mean value of all *reviewScore* values given by code authors in reverse reviewing stage.
- *Bonus score* plays an important role in this motivation model. It (*reviewBonus* in Fig. 2) may be positive or negative, depending on how close a code score is to the mean value of all code scores given by a review group. It is assigned by EduPCR4 automatically based on the motivation function defined as follows.

**Motivation Function**

Peer reviewers are asked to follow the clearly-defined source code score criteria listed in Appendix to write review comments and grade source code. These criteria are based on the best practices of software organizations such as GNU (http://www.gnu.org/prep/standards/standards.pdf), TIOBE (http://www.tiobe.com), and Parasoft (http://www.

parasoft.com/jsp/home.jsp) and are carefully selected such that new programmers can easily understand and apply when they write and review source code. A fundamental question in motivation model is how to determine the fairness of source code score given by a peer reviewer. The principle of the motivation model is that it should reward students who give fair code scores as well as penalize those who fail to give fair scores. Another requirement is that the motivation model can be easily implemented and executed by computer software without teachers' interference.

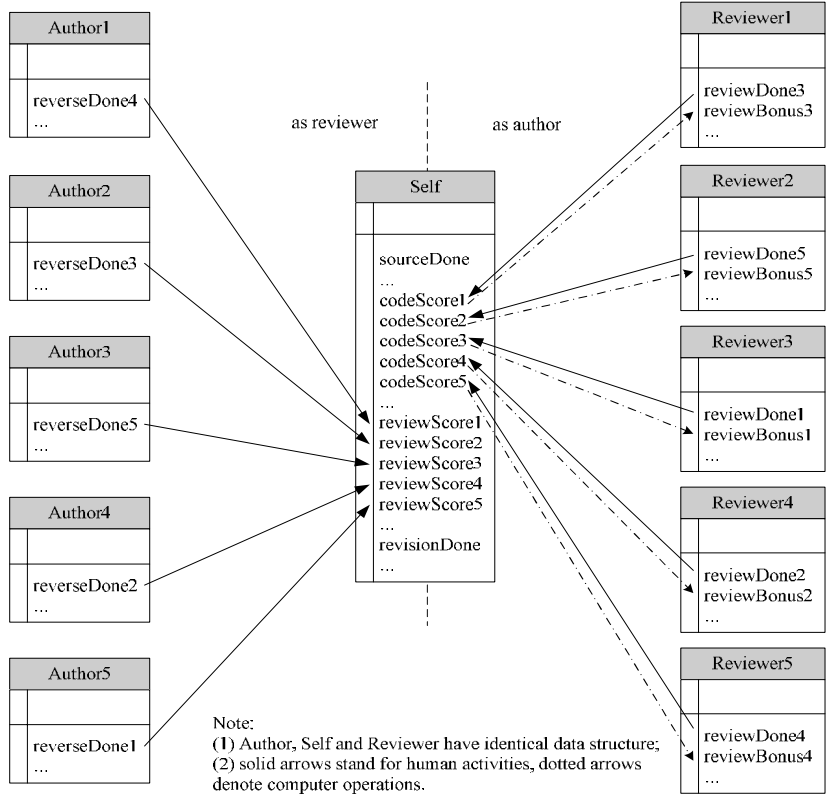

*Figure 2*. Data flow in a five-to-five peer review process

Lai and Lan (2006) aggregated the marks of multiple students to reduce personal bias through agent negotiation in peer assessment process. Sancho-Thomas et al. (2009) demonstrated that collective intelligence performs better than an individual intelligence. Thus, we use multiple peer reviewers to find a benchmark for its reward or penalty decision in the motivation model. A basic rule of the motivation model is that a student who gives a score close to the mean value of its review group scores should be awarded. On the other side, a student who gives a score far away from the mean value should be penalized. This model encourages students to mark peers' work as accurate as possible, i.e., to get consensus within one review group. Nonetheless, non-consensus is a common phenomenon in code review that has to be solved to make the motivation model robust. The solution will be discussed in the following sections.

Assume that the number of students in one class is $N$ and $x_i$ is the review quality score (*reviewScore*) of student $i$ in a review task, $0 \leq x_i \leq 100, 1 \leq i \leq N$. The motivation model adjusts $x_i$ to $x_i'$ using Formula 1.

$$x_i' = \begin{cases} x_i + A_1, D_i \leq T_1 \\ x_i + A_2, T_1 < D_i \leq T_2 \\ x_i - P_1, T_2 < D_i \leq T_3 \\ x_i - P_2, T_3 < D_i \end{cases} \quad (1)$$

In Formula 1, $D_i = |M - x_i|, M = \sum x_i / N$, and $A_1 > A_2 \geq 0$, $P_2 > P_1 \geq 0$, $1 < T_1 < T_2 < T_3$. $A$ and $P$ stand for *award* and *penalty* score respectively. $D$, $M$ and $T$ indicate *score deviation*, *mean value*, and *threshold* respectively. That is to say, this model breaks score deviations into four segments. Score deviation is the difference between a reviewer's code score and the group mean. Students whose score deviations fall into the former two segments will get positive incentive (award score). Similarly, students whose score deviations fall into the later two segments will receive negative incentive (penalty score).

The motivation model considers data dispersion before apply Formal 1. Specifically, the motivation model requires that the range of scores given to one specific source code should be greater than a predefined value. For example, in a specific source code submission, five reviewers (denoted A, B, C, D, and E), give a *codeScore* of 100, 99, 98, 97, 96 respectively. The data range, called *maxDiff*, is 4 (100 - 96). If we apply the above motivation formula without considering the small dispersion, student C will be awarded with $A_1$ scores, students B and D will get $A_2$ bonus, and student A and E will be penalized with $P_1$ scores. It is obviously unfair to all students since the *maxDiff* is comparatively small and there is a good consensus in this review group. Therefore, there is no need to penalize anyone if the value of *maxDiff* is less than a pre-defined value such as 10 or 15. An award may or may not be given to all reviewers when there is a good consensus. The values of *maxDiff*, $A_1$, $A_2$, $P_1$, $P_2$, $T_1$, $T_2$, and $T_3$ will be decided by the users of this motivation model based on their domain-specific requirements.

## Challenge and Solution

Even though it seems simple and reasonable, the proposed motivation model has a critical challenge. Non-consensus happens from time to time and may discount the performance of the motivation model. There are cases where non-consensus brings unfair penalties and inappropriate awards that both decrease participants' learning enthusiasm. This challenge has to be solved to make the motivation model practically useful.

Non-consensus originates from differences in different students' content knowledge or personal biases. It is a common phenomenon in peer assessment activity. In programming, different people have different or even opposite opinions on the same source code. In teaching programming language, it is preferable to let students to see the different opinions and learn from different views because most real world programming projects are team-based. Two common solutions to non-consensus are prevention and correction. Since prevention of non-consensus will be likely to increase the subjectivity of assessment results and suppress different opinions, the correction of non-consensus is a better choice for the non-consensus problem.

Non-consensus correction is to identify and correct any unfair awards or penalties given to code reviewers automatically by EduPCR4. In the proposed motivation model, the non-consensus may be generated only in code review stage because it is the only stage that multiple reviewers assess the same document. In this study, non-consensus results from one of the following three cases.
- *Case 1: Advanced program*. For example, when five reviewers are reviewing an "advanced" program written by a high-competence student, only one student can understand it and give it a high score. The other four reviewers give considerably low scores because they cannot understand it thoroughly. Thus, according to the motivation model, the student who gives higher score will be given a penalty.
- *Case 2: Low-competence reviewer*. For example, an author's program is quite understandable, but one of the reviewers cannot give a reasonable score because of his/her lack of content knowledge or programming skills. Meanwhile the other four reviewers give rational scores.
- *Case 3: Personal radicalness*. Some students tend to consistently give high score (or low score), not based on the code quality but their radical judgment standard or personal biases. In a worse scenario, this might be the result of irresponsibility.

In *Case 1*, both awarding and penalizing are unfair. The penalty to the high-competence reviewer is likely to hurt his/her learning motivation. To avoid this kind of negative motivation, a correction action is required. In this case, a teacher will inspect and make corrections to the scores of all reviewers. The award and deduction will be adjusted or even inverted. The adjustment, once received by all reviewers, gives positive motivation for not only the high competence students but also those students who do not appreciate the code in their initial reviews. In *Case 2*, the low-competence reviewer will be penalized according to the motivation model. The low-competence student is aware of the score penalty because the penalty is shown as a negative score adjustment to the student. It motives the student to read other review comments to improve his/her programming skills. In *Case 3*, the personal radicalness generates unfair assessment results. The system should identify it as soon as possible and warn students who give radical assessment. The warning helps a reviewer to change his/her scoring standards and align his/her assessment with the class assessment criteria.

EduPCR4 implements several algorithms to monitor non-consensus metrics and notify class teachers when it detects a large deviation or a significant radicalness metric. The implementation has the following components.

(1) *Deviation sorting*. The motivation model uses standard deviation (S.D.) to measure the score deviation of all members in one review group. It is assumed that the source code of a student $a$ (as an author) has been reviewed by $m_t$ reviewers in task $t$. The S.D. of the $m_t$ reviewers' score is defined in Formula 2, in which $e_{rat}$ stands for the score that the reviewer $r$ gives to the author $a$ in task $t$. $\overline{e_{at}}$ is the mean value of all scores given to author $a$ for task $t$. $Z_{at}$ measures the deviation of author $a$ in task $t$. The higher this value is, the more a non-consensus case is likely to happen. When the code review is done, all review groups are sorted by $Z_{at}$ in a descending order. It is worthing to point out that if there are only a few reviewers in a review group, i.e., $m_t$ is a small value such as *2* or *3*, the standard deviation, $Z_{at}$, will be biased. One way to reduce the small sample bias is to group review groups with similar mean scores to calculate the deviation. For example, all review groups with a mean score in the range of 85 to 90 are in a big estimation group to calculate each group's deviation. For small sample sizes where review groups have 4 or few reviewers, this will significantly reduce the estimation biases.

$$Z_{at} = \sqrt{\frac{1}{m_t}\sum_{r=1}^{m_t}(e_{rat}-\overline{e_{at}})^2}, \quad \overline{e_{at}} = \frac{1}{m_t}\sum_{r=1}^{m_t}e_{rat} \qquad (2)$$

(2) *Arbitration triggering*. If the S.D. value of one group reaches a given threshold, the teacher arbitration process will be triggered. The class teacher will receive a notice from EduPCR4 and be informed to deal with the arbitration as soon as possible. Teachers inspect the program, scores given to the program, and the reward/penalty scores to the reviewers. They have the right to modify reward and penalty scores of all members in the concerned groups. Because a teacher can use the deviation threshold to control the number of groups to be investigated, the arbitration workload is under the teacher's control.

(3) *Radicalness detection*. The radicalness of a reviewer $r$ is defined in Formula 3, in which $e_{rat}$ has the same meaning as in Formula 2. $m_t$ is the number of reviews conducted by reviewer $r$ in task $t$. $\overline{e_{rt}}$ is the mean value of all scores given by reviewer $r$ in task $t$. Basically $Z_r$ measures the deviance of all scores that reviewer $r$ gives to different authors in different tasks. A small $Z_r$ value means similar scores for different authors and different tasks: a sign of radicalness or personal bias. The less $Z_r$ value is, the more likely that the reviewer $r$ is radical. When a reviewer $r$ consistently gives similar scores to different authors in different tasks, EduPCR4 will detect a small $Z_r$ value. By the way, because $Z_r$ is an accumulative value, the more reviews have been completed by student $r$, the more accurate $Z_r$ is. $Z_r$ values of all students should be computed and compared after every assignment. When the $Z_r$ value of one student reaches a pre-defined threshold, the teacher warns the student to prevent him/her from giving radical reviews in future assignments. A teacher should let all students know that radical reviews will hurt their learning experience and weaken the performance of peer assessment.

$$Z_r = \sum_{t=1}^{T}\sum_{a=1}^{m_t}(e_{rat}-\overline{e_{rt}})^2 \Bigg/ \sum_{t=1}^{T}m_t, \quad \overline{e_{rt}} = \frac{1}{m_t}\sum_{a=1}^{m_t}e_{rat} \qquad (3)$$

## Empirical Evaluation

We developed the peer code review system EduPCR4 that implements the proposed motivation model and non-consensus detection algorithm. The empirical evaluation was conducted in a C programming language course in the fall semester of 2012 in a university. The participants were all 30 undergraduate students of the class. The code review criteria were delivered to them prior to the start of the course. We utilized a five-to-five reviewer assignment strategy that every student will review five peers' code and every source code is reviewed by five reviewers. Certainly, if the students regard five-to-five assignment strategy as a big workload and therefore have radical behavior, a four-to-four or three-to-three assignment strategy could be adopted. However, in this model based on peer assessment, the size of reviewers' assignment to one program should not be too small in order to keep fairness of peer assessment. Fortunately, our students did not realize the five-to-five reviewer assignment strategy was a big workload.

**Configuration of Critical Parameters**

Without losing generality, we assume that a task has a total score of 100. Because source code submission is the first event in the motivation model for an assignment, process score of this part was not included in our calculation formula. However, the revision submission provides reverse review and demonstrates student learning outcome, we give it 15 percent. Code quality and review quality takes up 30 percent equally. The process score of each *reviewDone* or *reverseDone* are 12.5 percent (totally 25).

After several attempts and fine tuning, we found good values of critical parameters in the motivation model. In order to maintain a normal score distribution and to prevent students from being too sensitive to awards and penalties, the portions of awards and penalties should be moderate. Thus, the values of $A_1$, $A_2$, $P_1$, and $P_2$ are set to 2, 0, 4, and 8 correspondingly. Moreover, $T_1$, $T_2$, and $T_3$ are set as 10%, 30%, and 60% of *maxDiff* respectively to make the bonus distribution comparatively reasonable.

**Identification of Non-consensus**

EduPCR4 implements the proposed non-consensus and radicalness detection algorithm. Table 2 and Table 3 are some sample standard deviation and radicalness data obtained from the class assignments. To protect students' privacy, their real ID numbers were coded in this study.

*Table 2.* Some review groups with highest standard deviation

| taskID | userID | S.D. |
|---|---|---|
| 1 | 6389 | 35.42 |
| 1 | 6394 | 35.14 |
| 1 | 6391 | 35.06 |
| 2 | 6380 | 36.83 |
| 2 | 6373 | 36.69 |
| 2 | 6390 | 32.29 |
| ... | ... | ... |

*Table 3.* The top 10 reviewers with highest radicalness

| userID | $Zr$ |
|---|---|
| 6369 | 9.54 |
| 6373 | 31.21 |
| 6389 | 53.05 |
| 6375 | 53.28 |
| 6390 | 77.27 |
| 6378 | 112.46 |
| 6370 | 119.85 |
| 6371 | 126.72 |
| 6384 | 128.25 |
| 6379 | 137.68 |

The review groups listed in Table 2 are *Case 1* (advanced program) and *Case 2* (low-competence reviewer) non-consensus. They triggered teacher arbitration process. *Case 1* is unreasonable non-consensus and *Case 2* is reasonable non-consensus. In the former case, the score deduction to the minority was unreasonable thus teachers had to manually correct the *reviewBonus* computed by EduPCR4. In the latter case, the score deduction to the minority of reviewers was reasonable and no further action required. Table 3 data is a *Case 3* non-consensus (personal radicalness). Teachers should talk to the students in first rows of Table 3 and warn them that their radical behavior in the assessment process would hurt the performance of themselves and the entire class. The sooner the communications were made in the course, the more reliable the peer assessment would be.

## Effectiveness of the Motivation Model

**Impacts on Students**

At the end of the class, students were asked to fill a simple questionnaire on this motivation model. The questionnaire results showed that a majority of participants like the award and penalty policies implemented in the motivation model.

Question: In source code review, the system awards a reviewer whose code review score is close to the code review group mean score. The system penalizes a reviewer whose code review score is far away from the code review group mean score. As a code reviewer, what do you think of this policy?
A. I like it. I will attend review training class and assess author's work carefully to get a bonus if possible.
B. I do not like it. I would rather evaluate an author's code according to my own understanding.
C. I do not care about award or penalty. I will mark author's work following my standard anyway.

The results proved that this motivation model of peer assessment was successful to some extend. This incentive policy did promote students' learning motivation. As a reviewer, about 78% of students had positive feedbacks (answer A) to this policy and would like to attend review training class and to assess author's work based on the class review criteria. About 22% of students had negative feedbacks (answer B) to this policy and would like to mark author's work according to their own understanding. There was no one who did not care about award or penalty (answer C).

**Impacts on Teachers**

The peer review process implemented in EduPCR4 significantly reduced the teacher's workload. In our experience, there were a total of 24 cases where teacher involvements (activities such as *arbitration*, *warning*, and *inspection*) were necessary. Most time-consuming work such as non-consensus identification had been done automatically by the peer assessment system EduPCR4. Each teacher-involved activity took about fifteen minutes on average. The teacher spent about six hours on the entire scoring process. In traditional teacher-grading process, the teacher's workload is high. Assume that it takes 5 minutes for a teacher to inspect and grade a student's program, the total time required will be sixty hours approximate (30 students, 12 assignment, 2 versions of each program). Therefore, the teachers can save about 90% of their time and efforts on grading.

Furthermore, with the automatic exception detection and manual correction mechanism such as arbitration, warning and inspection mentioned above, the teachers has little worry about the peer assessment quality. The peer assessment is reliable because every score is the mean of assessment results by up to five peers.

## Conclusion and Discussion

In this paper, a motivation model is proposed for a multiple-reviewer peer assessment process. The process is implemented in a peer code review information system named EduPCR4. The empirical study showed that the motivation model based on automatic exception detection and manual correction was practical and successful. Though the initial goal of the proposed motivation model was to release a teacher's burden of onerous quality control work that keeps the peer assessment fair and reliable, the survey showed that students obtained a feeling of achievement and conducted peer assessment more carefully and actively. Nevertheless, there are several topics to be investigated in the future research.

*Is there a bias generated from students trying to guess how others might feel so that they give an imaginary review rather than giving their honest opinion?* It is possible that some participants try to guess an average score in order to receive a bonus. However, we believe that the possibility of guess activities will be quite low with this peer assessment model. If a student does not collude with others, it is quite difficult to guess the average score of a particular program. On the contrary, a set of clearly-defined review criteria allows a student to assess a peer's work

honestly than guessing average score. Thus, the key issues are *simplicity*, *clarity*, *applicability*, and *usefulness* of review criteria as depicted in Appendix. A set of well-designed review criteria should be comprehensive, understandable and easy to follow. If well-designed marking criteria are easy to apply, why do students guess? In the long run, it is interesting to check the consistency of review comments and scores using linguistic analysis tool.

*How will teachers do if a majority of students give high scores to each other without prior agreement?* If every student tries to give high score to peers because he/she is afraid of getting penalty, even though they have no prior agreement, the peer assessment is not reliable. One possible solution to avoid this problem is to ask students to rank, instead of grade, a peer's work. For example, in each task, a reviewer is asked to rank, from 5 (best) to 1 (worst), all five programs he/she reviews. It alleviates the problem but not an ideal solution because it is unfair for programs with similar quality. Another possible solution is that the teacher manually reviews and grades a small sample of programs and compares the teacher's evaluation with peer assessment results. It helps to identify the problem but not solve it. To find a better solution to this problem is an interesting research topic that to be investigated in the future.

# Appendix: Example of evaluation criteria in peer assessment

**Part 1: Review criteria of assessing source code**

In this part, the default score is set to 100 points and a penalty strategy is utilized. If a student does not follow the coding rules, he or she will get some deduction points form the total score. In the worst case, when a student writes a program that does not compile or executes as specified, he/she may get 0 score.

*(1) Code comments*

*File header comment*. The recommended file header comment is as follows. Missed or bad-written file header comment may get a deduction of 2 through 10 points.

```
/**
 * This program is to demonstrate something...
 * @author  : Your Name (yourname@yourorg.com)
 * @date    : May 20, 2013
 * @version : 1.0.0
 */
```

*Block comment*. Every block should have a comment at the top of the block and a blank line above the comment. The violation to this rule may get a deduction of 2 through 10 points.

*Line comment*. The key statements such as loop control statement or important algorithm implementation should have line comment at the end of corresponding lines. The violation to this rule may get a deduction of 2 through 10 points.

*(2) Identifier naming*

The violation to the following rules may get a deduction of 1 through 20 points.
- Constant should be written in the format of NO_OF_STUDENTS. That is to say, all characters should be in capital and words should be separated with underscore mark.
- Use *Camel* and *Pascal* naming conventions instead of some vanishing naming conventions such as *Hungarian* convention, *Positional* notation, etc. (see http://en.wikipedia.org/wiki/Naming_convention_(programming))
- Avoid too simple naming such as *i*, *jj*, *kkk*, etc.
- In Java programming, try to hide the properties of object and define getter and setter methods in class to manipulate them. The method names should be like *getSome*() or *setThing*().

*(3) Improper or poor layout*

The violation to the following rules may get a deduction of 1 through 20 points.
- Use Java indent style as follows:
```
int Foo(bool isBar) {
    if (isBar) {
        bar();
        return 1;
    } else {
        return 0;
    }
}
```

- Use white space instead of tab key because the program may look different in different editors.

Hint: some "intelligent" editors such as *Netbeans* and *Eclipse* have the function to format codes automatically. If you cannot manage the code layout by yourself, take full advantage of this feature to avoid losing points.

### *(4) Program does not meet the assignment requirement*

- The objective of every program is able to run and solve a problem.
- If you feel intuitively that an author's program cannot execute properly and you have no time to test it in your machine, you can give a deduction of 15 to 25 points.
- If you have proved that an author's program cannot execute properly by testing, you can give a deduction of 20 to 40 points.

**Part 2: Reverse review criteria (reverse assessment)**

In this part, the default score is set to 0 point and an incremental strategy is applied. As an author, you assess a reviewer's work item by item and add them up to get a final score. Note: There is no award or penalty on reviewer assessment, please assess reviewer's work honestly.

### *(1) Comments are complete.*

If the reviewer have made all (on your own feeling) advices or suggestions, you can give up to 25 points to that reviewer. Otherwise, if few advices or recommendations are written and you find many shortcomings in your source code when you revise your code or read other reviewers' comments, you can give it a low score.

### *(2) Comments are constructive.*

If a reviewer gives constructive comments that are helpful for you to improve your code, you can give the reviewer up to 25 points. Otherwise, for less helpful or useless comments, even though there are many of them, you give it a low score.

### *(3) Comments are objective and fair.*

If a reviewer gives objective and fair comments, you can give him/her up to 25 points. Otherwise, you give a low score.

### *(4) Comments are concise and clear.*

If a reviewer states comments clearly and concisely, i.e., the comments are quite understandable, you can give him/her up to 25 points also.